# Framework Model for Database Replication within the Availability Zones

Ala'a Atallah A. Al-Mughrabi[1]     Hussein H. Owaied[2]

[1] Computer Information System Department, Middle East University, Amman, Jordan
almughrabi2004@yahoo.com

[2] Computer Science Department, Middle East University, Amman, Jordan
Owaied@yahoo.com

**Abstract**

This paper presents a proposed model for database replication model in private cloud availability regions, which is an enhancement of the SQL Server AlwaysOn Layers of Protection Model presents by Microsoft in 2012. The enhancement concentrates in the database replication for private cloud availability regions through the use of primary and secondary servers. The processes of proposed model during the client send Write/Read Request to the server, in synchronous and semi synchronous replication level has been described in details also the processes of proposed model when the client send Write/Read Request to the Primary Server presented in details. All the types of automatic failover situations are presented in this thesis. Using the proposed models will increase the performance because each one of the secondary servers will open for Read / Write and allow the clients to connect to the nearby secondary and less loading on each server.

Keywords: *Availability Regions, Cloud Computing, Database Replication, SQL Server AlwaysOn, Synchronization.*

## 1. Introduction

[1] Shipsey 2010 describes one of the major components of an IT infrastructure and claims that, the organizations generally cannot afford to simply throw out all of their old technology and replace it. It is cheaper (and involves less staff training) to use middleware and other technologies which integrate the old and the new. Rather than buying all of the hardware necessary to support websites, intranets and extranets, many smaller companies choose to use web hosting services instead. Cloud computing provides the hardware, software, expertise and security necessary for a company to have a web presence without becoming a major distraction (in terms of time and money) from the core business.

Hitachi Data Systems in 2007 said that main problem to synchronous replication is its distance limitation. Fiber Channel, the primary enterprise storage transport protocol, can theoretically spread up to 200 kilometers (km) or 124 miles. But, latency quickly becomes a difficult as propagation delays increase with increased distance [2].

Ashok and Randal in 2008) show how the transactional replication uses the Transactional Replication Architecture to move changes between the servers in a replication topology [3].

Service management of cloud services become as critical as indicated in the next decade need to look at the maturity status and evolution of service management tools related to the cloud computing. While many clients consider their existing tool landscape as slightly better equipped to cope with the new challenges ahead. So the need to have better automation, control and visualization service management. Service security and the ability to control data/information access and usage is a must today and will be followed by an increased focus on identity management in future [4].

Cloud computing is a model that enables, for everywhere, convenient, and on-demand, network access to a shared pool of configurable computing resources [5]. Newton in 2010 described how the data availability is considered and has a strong related with cloud computing since the data the business need to complete work available   is on single cloud so data availability becomes a great challenge [6].

Data replication is a key technology in distributed systems that enable higher availability and performance [7]. The main goal of the Data replication is maintaining multiple copies of data, called replicas, on separate servers at different locations [8]. Data replication it's an important enabling technology for distributed services [9]. Replication improves availability by allowing access to the data even when some of the replicas are unavailable [10].

## 2. Proposed Framework Model

Usually, most of availability zones covered by any provider have been regarded as synchronized database replication within 100 km between physical data center that connected with each other with a secure fast link media otherwise the replication will be asynchronies. Therefore the database replication in the Private Cloud within 100 km between the master server and slave server of the provider has no problems. Therefore physical data center connected with each other with a secure fast link media to all availability zones in the same region.

Fig.1 presents an overview of the proposed model deals with the availability zones, covered by the provider, greater than 100 km. The reason for chosen such availability zones, is usually the database replication is asynchronies in such availability zones, so the latency time is greater than the latency time in the availability zones within 100 km.

In this new model there is now direct connection from primary server to the servers in the semi synchronous replication level. Unless the secondary in the synchronous replication level is failed so the connection will be handled as the current model.

### 2.1 General Architecture of the Read Request Model

Fig.2 presents the General Architecture of the Read Request Model which shows the request can be to the either primary, Secondary in synchronous replication level or Secondary in semi synchronous replication level. Since the each site in the whole region using the proposed model has a copy of the database with current information so the clients can read the data that's needed from any nearby sever in the whole region and that allowing to high performance of the cloud application and satisfaction the end users.

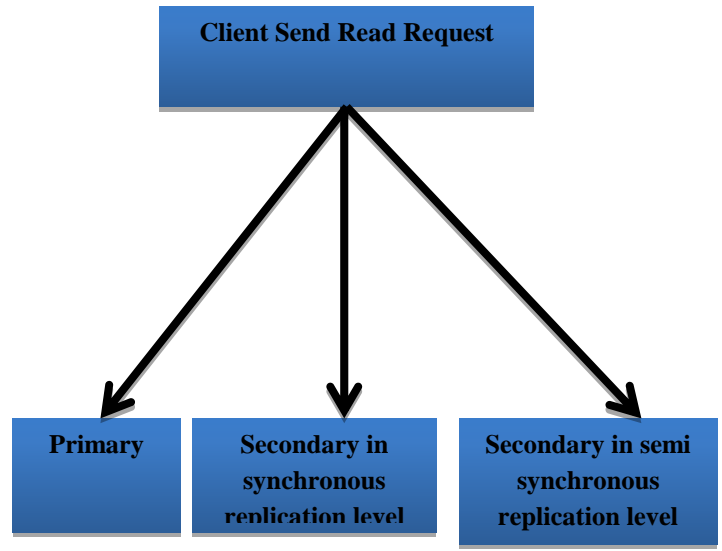

Fig.2 General Architecture of the Read Request Model

### 2.2 General Architecture of the Proposed Write Request Model

Fig.3 presents the General Architecture of the write request model which shows the request can be to the either primary, Secondary in synchronous replication level or Secondary in semi synchronous replication level.

Using the proposed model the write request will be sending to the all server in the region and this increase the performance of the network because the write request is distributed with the whole servers in the region Instead in just allow the write request to be apply to the primary server.

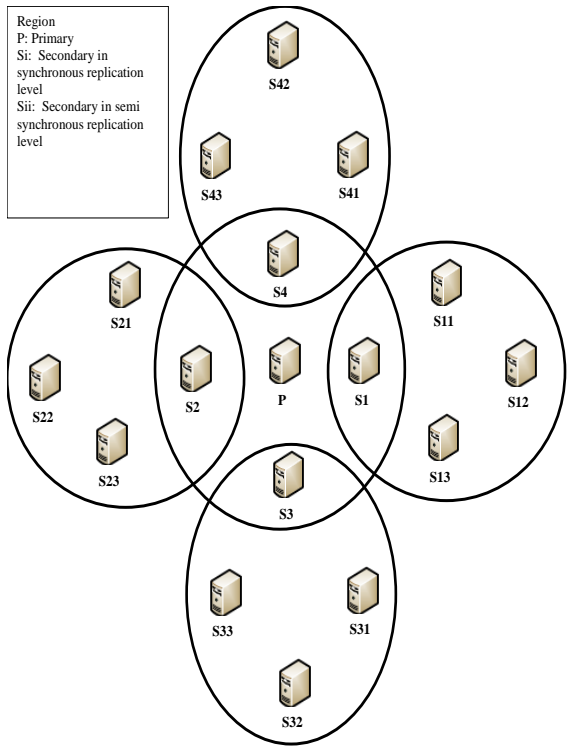

Fig.1 an overview of proposed model deals with the availability zones

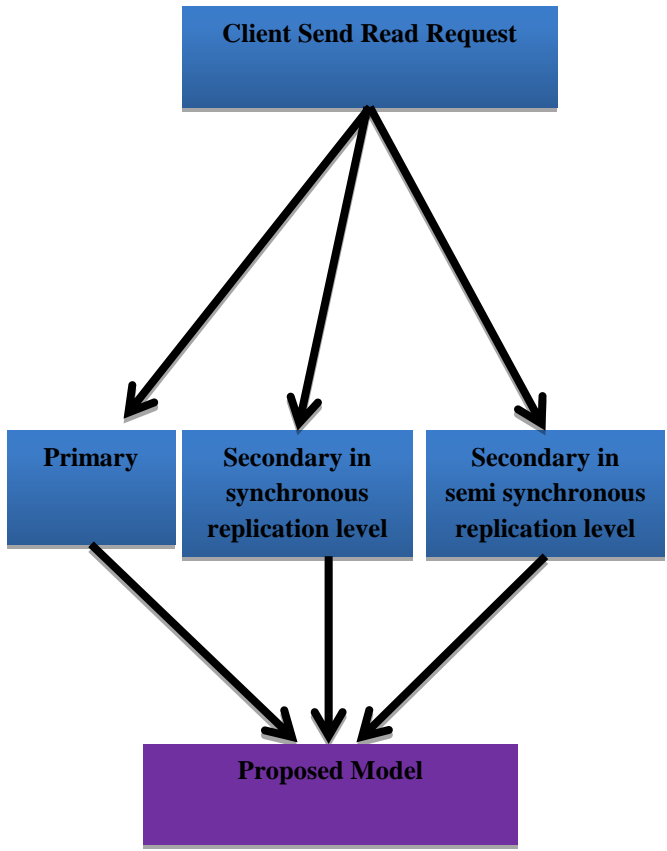

Fig.3: General Architecture of the Proposed Write Request Model

### 2.3 Assumptions for Applying Proposed Model

1. Put the primary server in the center of the region as possible as you can. In this point while the primary is in the center of the region so you can expand the region with the new model in the four direction of the primary with the same performance without effecting one direction on the others.
2. Add four secondary servers to be replication servers with synchronous distance in the four directions from the primary server. In this point you will put four secondary as synchronous replication while there is at maximum two secondary with synchronous mode in the SQL 2012 replication model, you will call this synchronous replication level.
   Also in the SQL 2012 replication model you can failover automatically to one from the two synchronous secondary servers. But with using the new model you can failover to four secondary synchronous servers, as every synchronous secondary server will have priority number to failover on it so if the primary fail you will automatically transfer to the secondary server that have the priority number one, if this new primary fail you will transfer to the secondary server have priority number two and so one. With using this feature we will increase the availability to the maximum.
3. Add three other secondary servers with synchronous distance from each one of the main four secondary servers on the synchronous replication level. We will call these new secondary servers semi-synchronous replication Level. With using this new feature we can expand the region to the double normal size with using the traditional.
4. The most important note that using the proposed model each secondary servers will open for both read / Write. Using the proposed models will increase the performance because each one of the secondary servers will open for Read / Write and allow the clients to connect to the nearby secondary and less loading on each server because read and write request will be distributed between all servers in the entire region.

### 2.4 The processes of write request in semi synchronous replication level

Fig. 4 presents the summary of the processes for the proposed model during the Client Send Write Request to the server in semi synchronous replication level. The more details descriptions of these processes will be described according to the following phases:

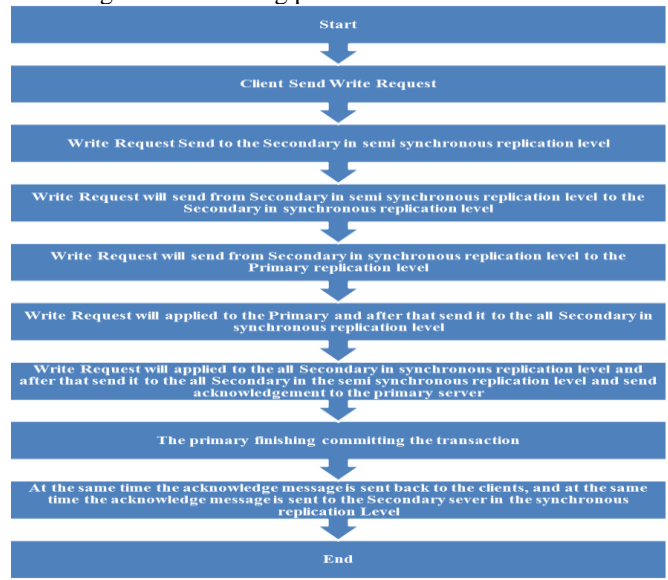

Fig.4: processes of write request in semi synchronous replication level

Phase one: client send write request to the secondary in semi synchronous replication level

Fig.5 presents the processes in details that happened in the secondary in semi synchronous replication level when secondary in semi synchronous replication level receiving the write request from client until passing the write request to the Secondary in synchronous replication level, as shown in the pseudo code in fig.6 .

Begin
Input Client_Write_Request
Input Secondary_Write_Log_sequence_number
Input Primary_Write_log_sequence_number
Input Status
If server= secondary in semi synchronous replication and
Client_Write_Request<> Nothing and
Primary_Write_log_sequence_number=Nothing
Secondary_Write_Log_sequence_number=Nothing and
Status=Nothing Then
/*Determine_Status is a Function that determine status of the transaction for each phase*/
Set Status= Determine_Status (Client_Write_Request)
/*Compute_Secondary_Write_Log_sequence_number (Client_Write_Request) is a Function that create Secondary Write Log sequence number */
Set Secondary_Write_Log_sequence_number=Compute_Secondary_Write_Log_sequence_number (Client_Write_Request)
/* Send_write_request_to_secondary_in_synchronous_replication_level () is a Function redirect the transaction log to the secondary in synchronous replication level */
Send_write_request_to_secondary_in_synchronous_replication_level ()
End if
Outputs:
Client_Write_Request
Secondary_Write_Log_sequence_number=S42W0001
Status= pending from Secondary in semi synchronous replication level
Primary_Write_log_sequence_number=Nothing
End

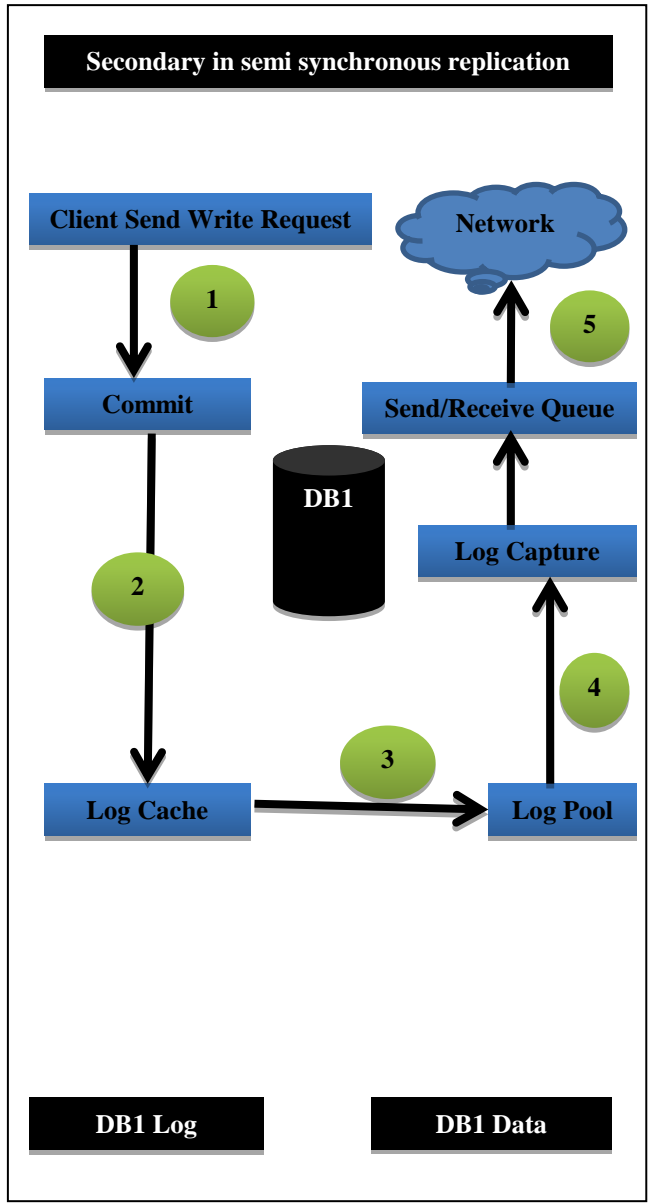

Fig.5: Secondary in semi synchronous replication level Steps (1-5)

Fig.6 Pseudo code for phase one

Phase two: secondary in synchronous replication level receive write request from secondary in semi synchronous replication level

Fig.7 presents the processes in details that happened in the secondary in synchronous replication level when receiving the write request from secondary in semi synchronous replication level until passing the write request to the primary serve, as shown in the pseudo code in Fig.8.

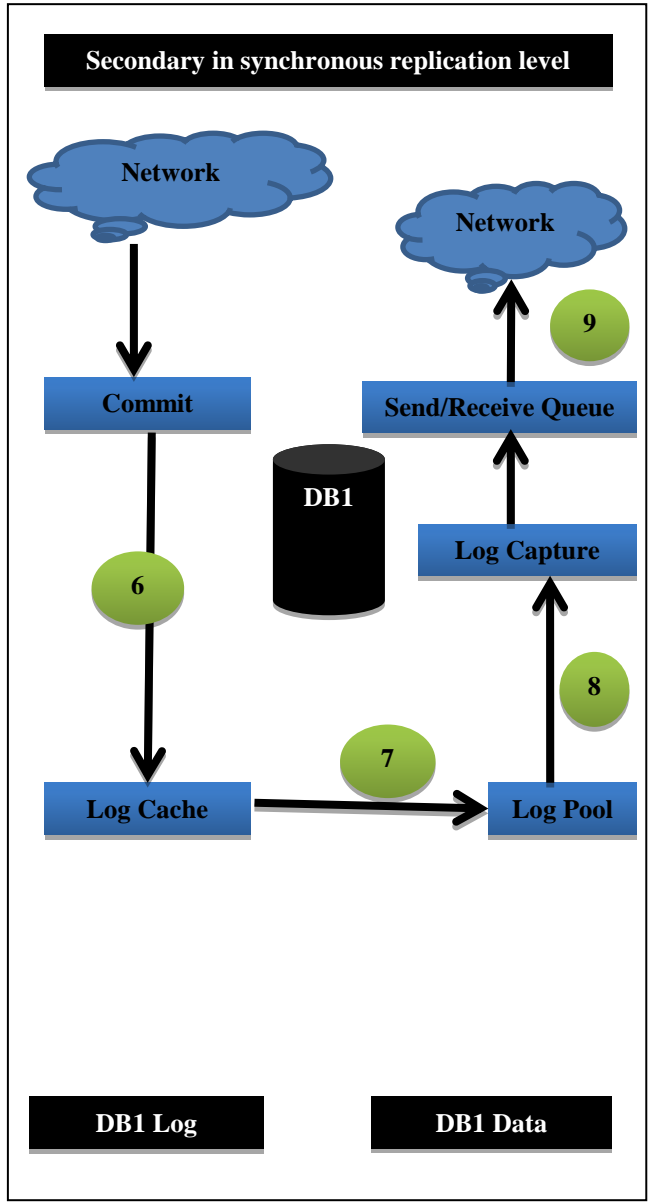

Fig.7: Secondary in synchronous replication level Steps (6-9)

Begin
Input:
Client_Write_Request
Secondary_Write_Log_sequence_number=S42W0001
Status= pending from Secondary in semi synchronous replication level
Primary_Write_log_sequence_number=Nothing
If server= secondary in synchronous replication and
Client_Write_Request<> Nothing and
Primary_Write_log_sequence_number=Nothing
Secondary_Write_Log_sequence_number= S42W0001and
Status= pending from Secondary in synchronous replication level Then
/*Determine_Status is a Function that determine status of the transaction for each phase*/
Set Status= Determine_Status (Client_Write_Request)
/* Send_write_request_to_primary () is a Function redirect the transaction log to the primary server */
Send_write_request_to_primary ()
End if
Outputs:
Client_Write_Request
Secondary_Write_Log_sequence_number=S42W0001
Status= pending from Secondary in synchronous replication level
Primary_Write_log_sequence_number=Nothing
End

Fig.8 Pseudo code for phase two (secondary in synchronous replication level receive write request from secondary in semi synchronous replication level)

**Phase three: primary server receive write request from secondary in semi synchronous replication level**

Fig. 9 presents the processes in details that happened in the primary server when primary server receive the write request from secondary in synchronous replication level until passing the write request to the in all secondary servers in synchronous replication level , as shown in the pseudo code in Fig.10.

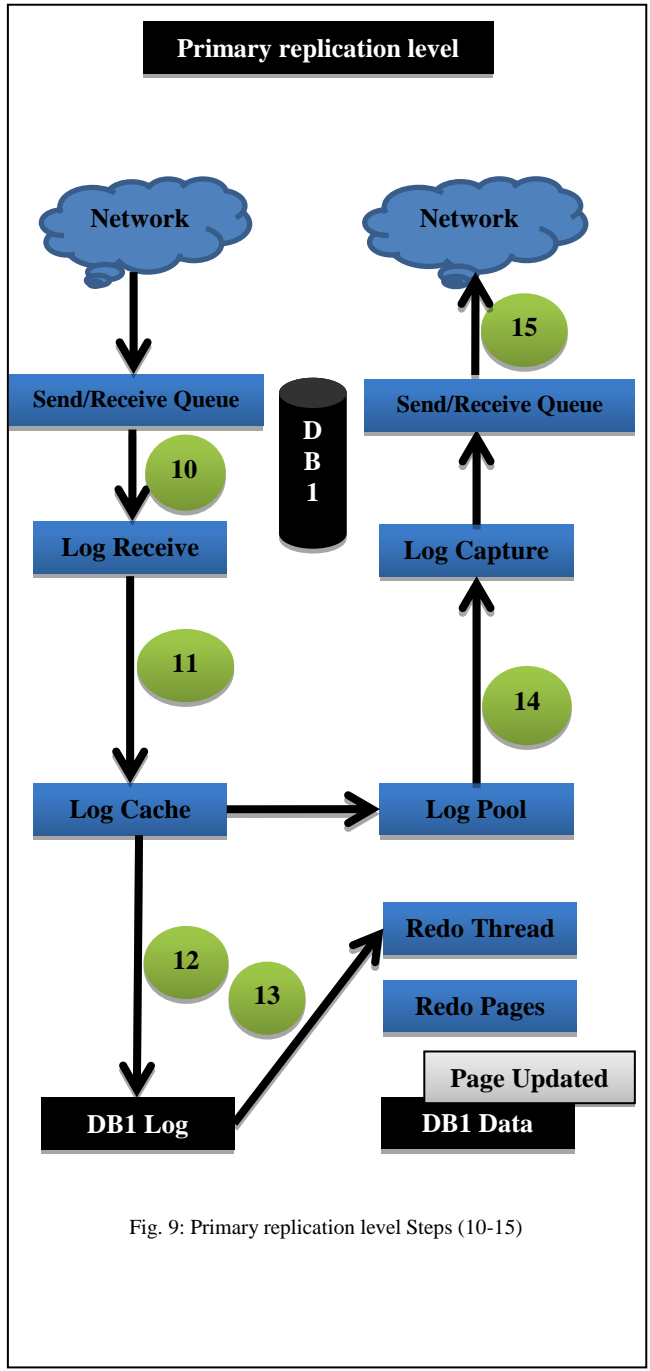

Fig. 9: Primary replication level Steps (10-15)

```
Begin
Input:
Client_Write_Request
Secondary_Write_Log_sequence_number=S42W0001
Status= pending from Secondary in synchronous replication level
Primary_Write_log_sequence_number=Nothing
If server= primary and
Client_Write_Request<> Nothing and
Primary_Write_log_sequence_number=Nothing
Secondary_Write_Log_sequence_number= S42W0001 and
Status= pending from Secondary in synchronous replication level Then
//write the log to the database log and database data
Applying_Write_Request ()
/*Determine_Status is a Function that determine status of the transaction for each phase*/
Set Status= Determine_Status (Client_Write_Request)
/*Compute_Primary_Write_Log_sequence_number (Client_Write_Request) is a Function that create primary Write Log sequence number */
Primary_Write_Log_sequence_number=Compute_Primary_Write_Log_sequence_number (Client_Write_Request)
/* Send_write_request_to_all_secondary_in_synchronous_replication_level () is a redirect the transaction log to the all secondary in synchronous replication level */
Send_write_request_to_all_secondary_in_synchronous_replication_level ()
End if
Outputs:
Client_Write_Request
Secondary_Write_Log_sequence_number=S42W0001
Status= Primary commit
Primary_Write_log_sequence_number=P01W0001
End
```

Fig.10 Pseudo code for phase three (primary server receive write request from secondary in semi synchronous replication level)

Phase four: Secondary in synchronous replication level receive commit write request from primary server

Fig.11 presents the processes in details that happened in the Secondary in synchronous replication when Secondary in synchronous replication receive the commit write request from the primary server until passing the commit write request to the in all secondary servers in semi synchronous replication level , as shown in the pseudo code in Fig. 12.

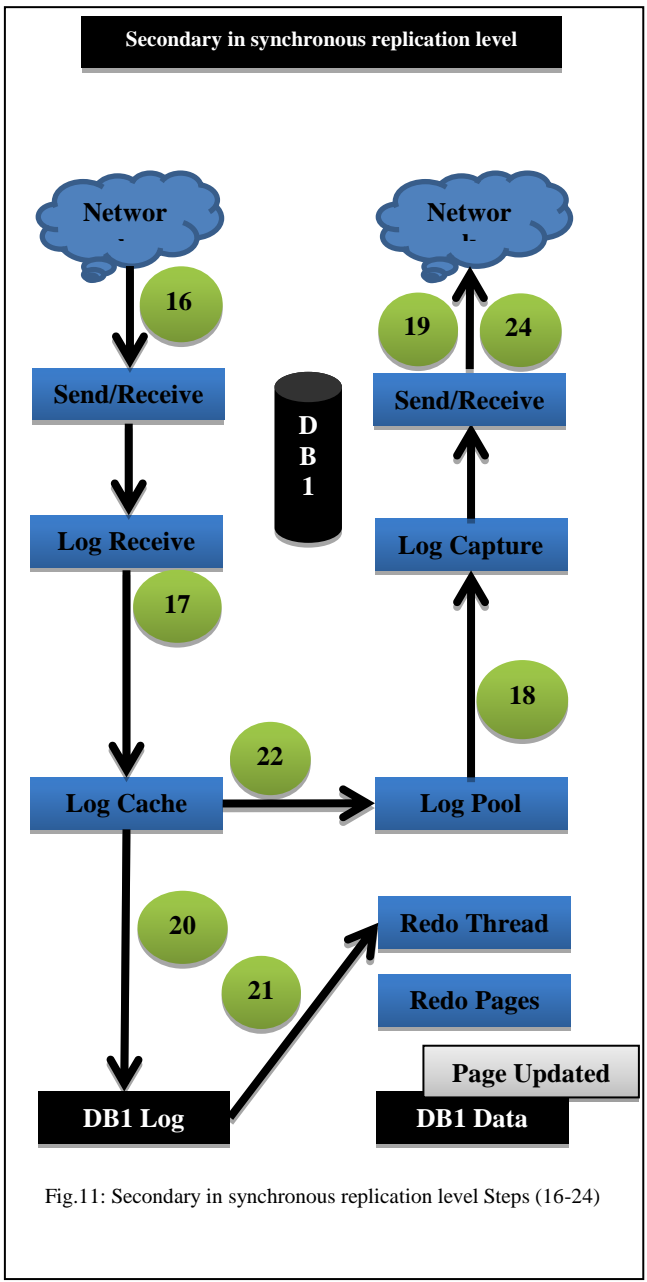

Fig.11: Secondary in synchronous replication level Steps (16-24)

```
Begin
Input:
Client_Write_Request
Secondary_Write_Log_sequence_number=S42W0001
Status= primary commit
Primary_Write_log_sequence_number= P01W0001
If server= secondary in synchronous replication and
Client_Write_Request<> Nothing and
Primary_Write_log_sequence_number<>Nothing
Secondary_Write_Log_sequence_number<> Nothing and
Status= primary commit Then
//write the log to the database log and database data
Applying_Write_Request ()
/*Determine_Status is a Function that determine status of the transaction for each phase*/
Set Status= Determine_Status (Client_Write_Request)
/*
Send_write_request_to_all_secondary_in_semi_synchronous_replication_level () is a function that redirect the transaction log to the secondary in semi synchronous replication level*/
Send_write_request_to_all_secondary_in_semi_synchronous_replication_level ()
/* Send_Staus_To_Primary () is a function that redirect the transaction log to the primary server*/
Send_Staus_To_Primary ()
End if
Outputs:
Client_Write_Request
Secondary_Write_Log_sequence_number=S42W0001
Status= acknowledgement from synchronous replication Level
Primary_Write_log_sequence_number=P01W0001
End
```

Fig.12 Pseudo code for phase four (Secondary in synchronous replication level receive commit write request from primary server)

**Phase five: the primary receive acknowledgement the write request from Secondary in synchronous replication level**

Fig.13 presents the processes in details that happened in the primary after acknowledgement is come back from the Secondary in synchronous replication level, as shown in the pseudo code in Fig.14.

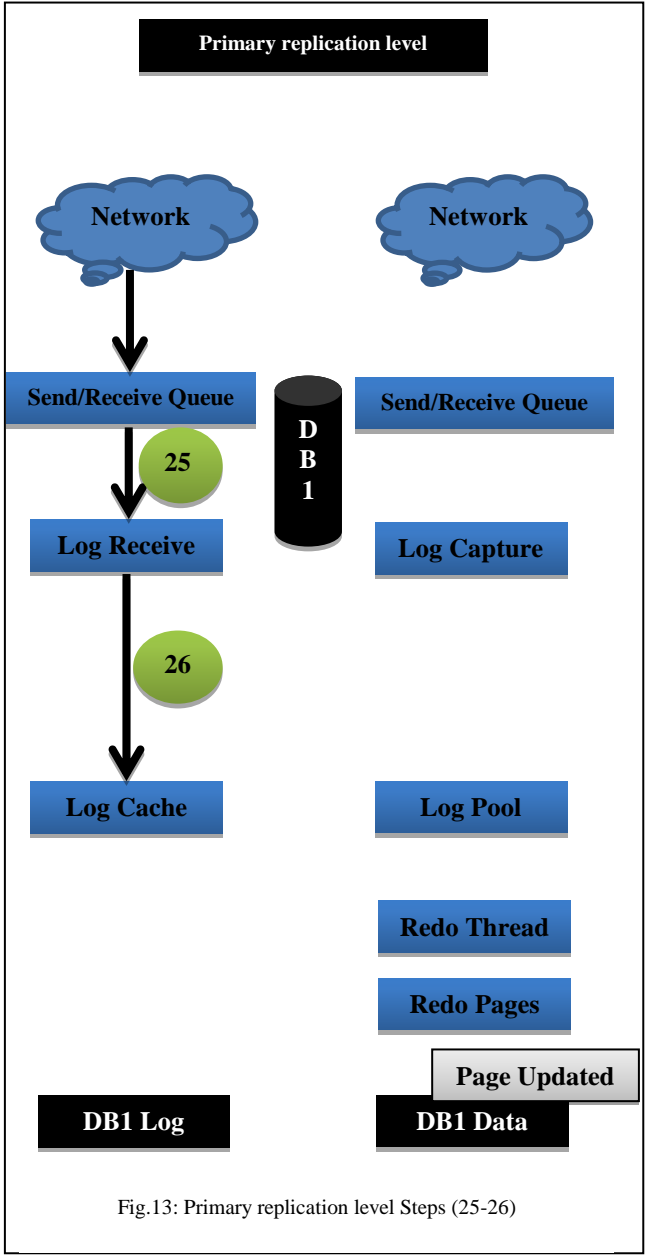

Fig.13: Primary replication level Steps (25-26)

```
Begin
Input:
Client_Write_Request
Secondary_Write_Log_sequence_number=S42W0001
Status= acknowledgement from synchronous replication Level
Primary_Write_log_sequence_number= P01W0001
If server= primary and
Client_Write_Request<> Nothing and
Primary_Write_log_sequence_number<>Nothing
Secondary_Write_Log_sequence_number<>   Nothing and
Status= acknowledgement from synchronous replication Level Then
//The Primary finishes committing the data
Primary_Finish_Commit_Write_Request ()
End if
End
```

Fig. 14 Pseudo code for phase five (the primary receive acknowledgement the write request from Secondary in synchronous replication level)

Phase six: the Secondary in semi synchronous replication level receive primary commit write request from Secondary in synchronous replication level

    Fig.15 presents the processes in details that happened in the Secondary in semi synchronous replication level after primary commit is come back from the Secondary in synchronous replication level, as shown in the pseudo code in Fig.16.

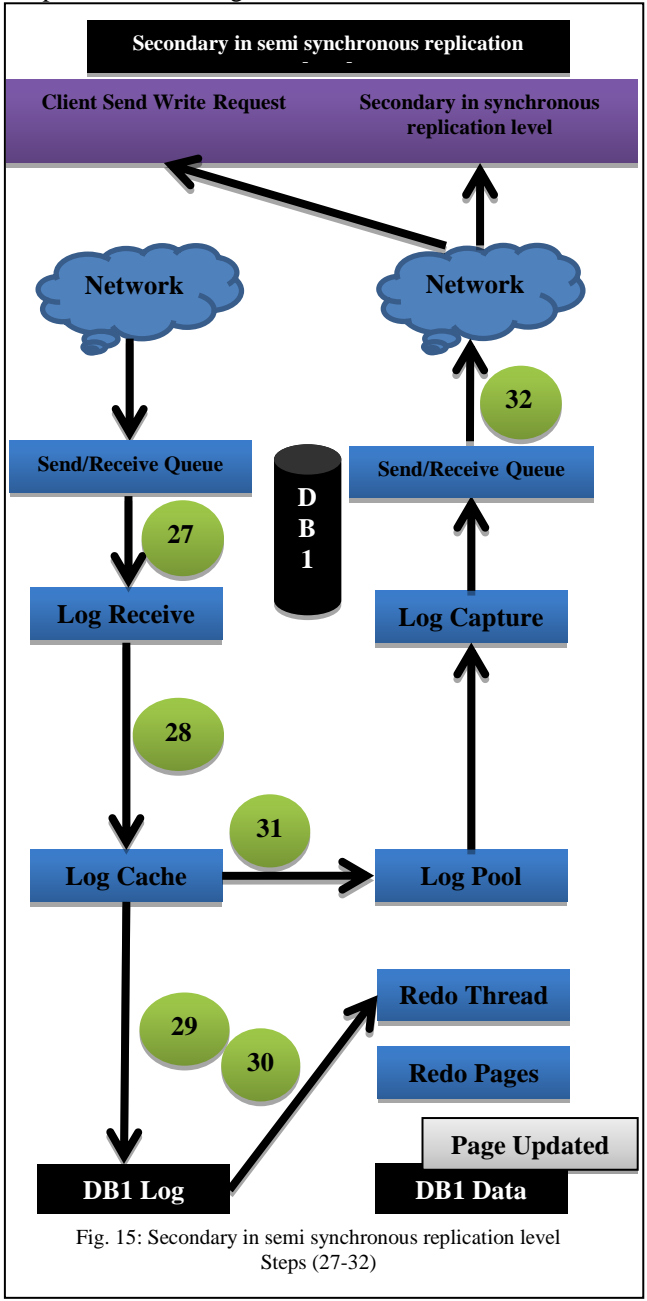

Fig. 15: Secondary in semi synchronous replication level Steps (27-32)

Begin

Input:

Client_Write_Request

Secondary_Write_Log_sequence_number=S42W0001

Status= primary commit

Primary_Write_log_sequence_number= P01W0001

If server= secondary in semi synchronous replication and

Client_Write_Request<> Nothing and

Primary_Write_log_sequence_number<>Nothing

Secondary_Write_Log_sequence_number<> Nothing and

Status= primary commit Then

//write the log to the database log and database data

Applying_Write_Request ()

/*Determine_Status is a Function that determine status of the transaction for each phase*/

Set Status= Determine_Status (Client_Write_Request)

/* Send_Staus_acknowledgement_synchronous_replication_Level () is a function that send acknowledgement to the secondary server in synchronous replication Level */

Send_Staus_acknowledgement_synchronous_replication_Level ()

End if

Outputs:

Client_Write_Request

Secondary_Write_Log_sequence_number=S42W0001

Status= acknowledgement from semi synchronous replication Level

Primary_Write_log_sequence_number=P01W0001

Display= acknowledge message is sent back to the clients

End

Fig. 16 Pseudo code for phase six (Phase the Secondary in semi synchronous replication level receive primary commit write request from Secondary in synchronous replication level)

## 2.5 Write Request Processes for Primary server

Fig.17 presents the summary of the processes for the proposed model during the Client Send Write Request to the primary server. The more details descriptions of these processes will be described according to the following phases:

Phase one: client send write request to the primary server

Fig.18 presents the processes in details that happened in the primary when the primary server receive the write request from client until passing the write request to the Secondary in synchronous replication level , as shown in the pseudo code in Fig.19.

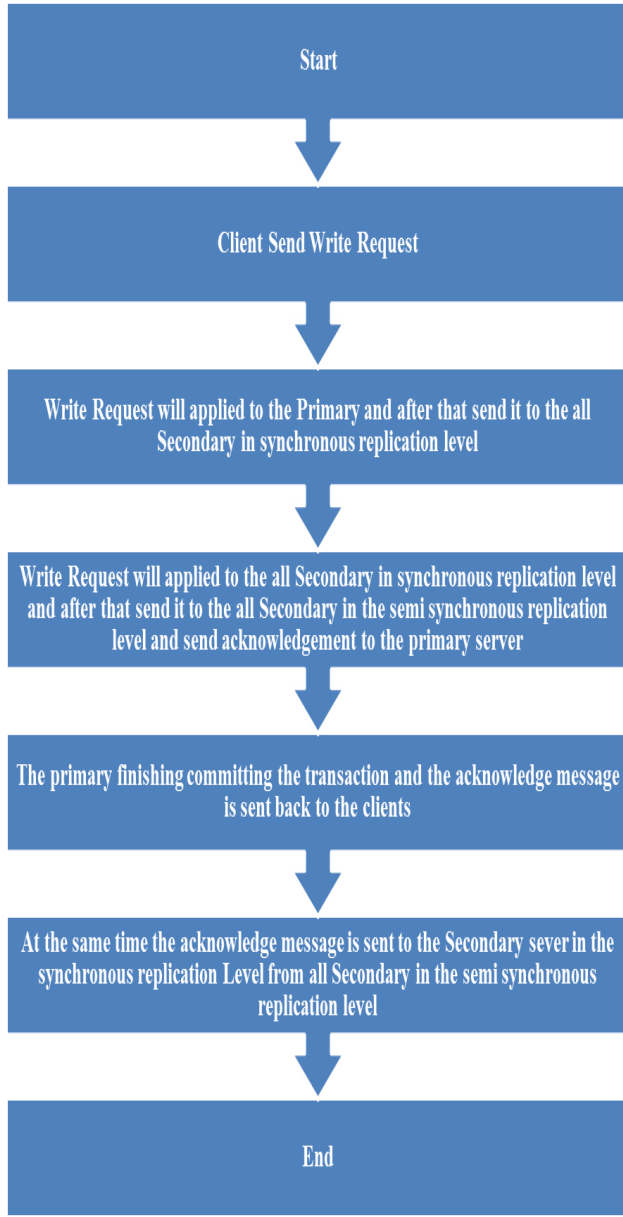

Fig.17: Flowchart for the processes of write request from the Primary server

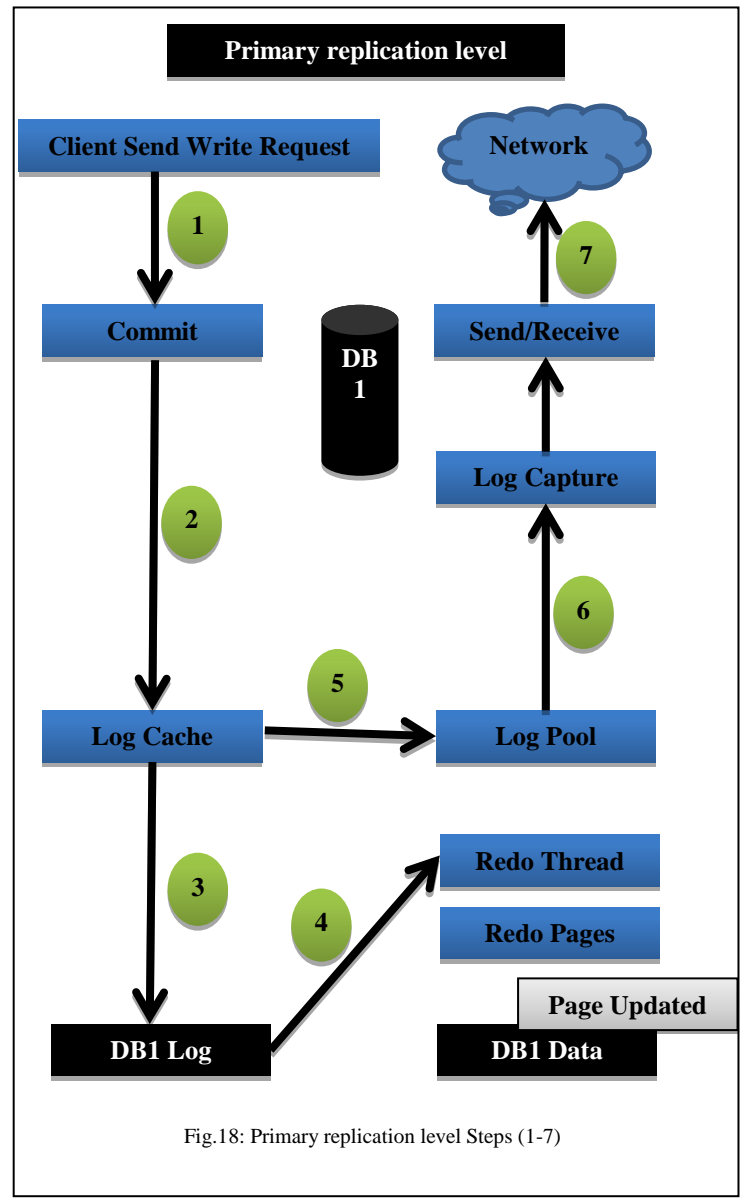

Fig.18: Primary replication level Steps (1-7)

```
Begin
Input:
Client_Write_Request
Secondary_Write_Log_sequence_number=Nothing
Status=Nothing
Primary_Write_log_sequence_number=Nothing
If server= primary and
Client_Write_Request<> Nothing and
Primary_Write_log_sequence_number=Nothing
Secondary_Write_Log_sequence_number=Nothing and
Status= Nothing Then
//write the log to the database log and database data
Applying_Write_Request ()
/*Determine_Status is a Function that determine status of the transaction for each phase*/
Set Status= Determine_Status (Client_Write_Request)
/*Compute_Primary_Write_Log_sequence_number
(Client_Write_Request) is a Function that create primary Write Log sequence number */
Primary_Write_Log_sequence_number=Compute_Primary_Write_Log_sequence_number (Client_Write_Request)
/*Send_write_request_to_all_secondary_in_synchronous_replication
_level () is a redirect the transaction log to the all secondary in synchronous replication level */
Send_write_request_to_all_secondary_in_synchronous_replication_level ()
End if
Outputs:
Client_Write_Request
Secondary_Write_Log_sequence_number=Nothing
Status= Primary commit
Primary_Write_log_sequence_number=P01W0001
End
```

Fig.19 Pseudo code for phase one (client send write request to the primary server)

Phase two: Secondary in synchronous replication level receive commit write request from primary server without Secondary write log sequence number:

Fig.20 presents the processes in details that happened in the Secondary in synchronous replication when Secondary in synchronous replication receive the commit write request from the primary server until passing the commit write request to the in all secondary servers in semi synchronous replication level without Secondary write log sequence number, as shown in the pseudo code in Fig.21.

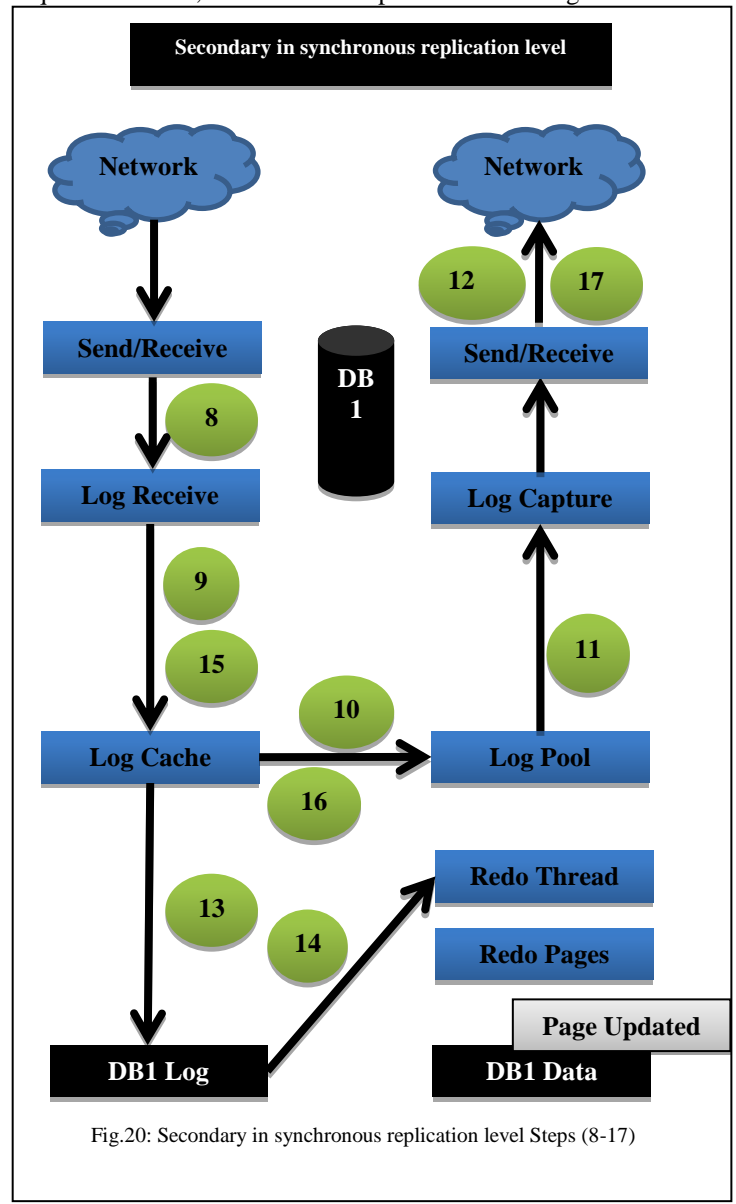

Fig.20: Secondary in synchronous replication level Steps (8-17)

Begin

Input:

Client_Write_Request

Secondary_Write_Log_sequence_number=Nothing

Status= primary commit

Primary_Write_log_sequence_number= P01W0001

If server= secondary in synchronous replication and

Client_Write_Request<> Nothing and

Primary_Write_log_sequence_number<>Nothing

Secondary_Write_Log_sequence_number=Nothing and

Status= primary commit Then

//write the log to the database log and database data

Applying_Write_Request ()

/*Determine_Status is a Function that determine status of the transaction for each phase*/

Set Status= Determine_Status (Client_Write_Request)

/*Send_write_request_to_all_secondary_in_semi_synchronous_replication_level () is a function that redirect the transaction log to the secondary in semi synchronous replication level*/

Send_write_request_to_all_secondary_in_semi_synchronous_replication_level ()

/* Send_Staus_To_Primary () is a function that redirect the transaction log to the primary server*/

Send_Staus_To_Primary ()

End if

Outputs:

Client_Write_Request

Secondary_Write_Log_sequence_number=Nothing

Status= acknowledgement from synchronous replication Level

Primary_Write_log_sequence_number=P01W0001

End

Fig.21 Pseudo code for phase two (Secondary in synchronous replication level receive commit write request from primary server without Secondary write log sequence number)

Phase three: the primary receive acknowledgement the write request from Secondary in synchronous replication level and send acknowledge to client

Fig.22 presents the processes in details that happened in the primary after acknowledgement is come back from the Secondary in synchronous replication level and acknowledge send back to the client, as shown in the pseudo code in Fig.23.

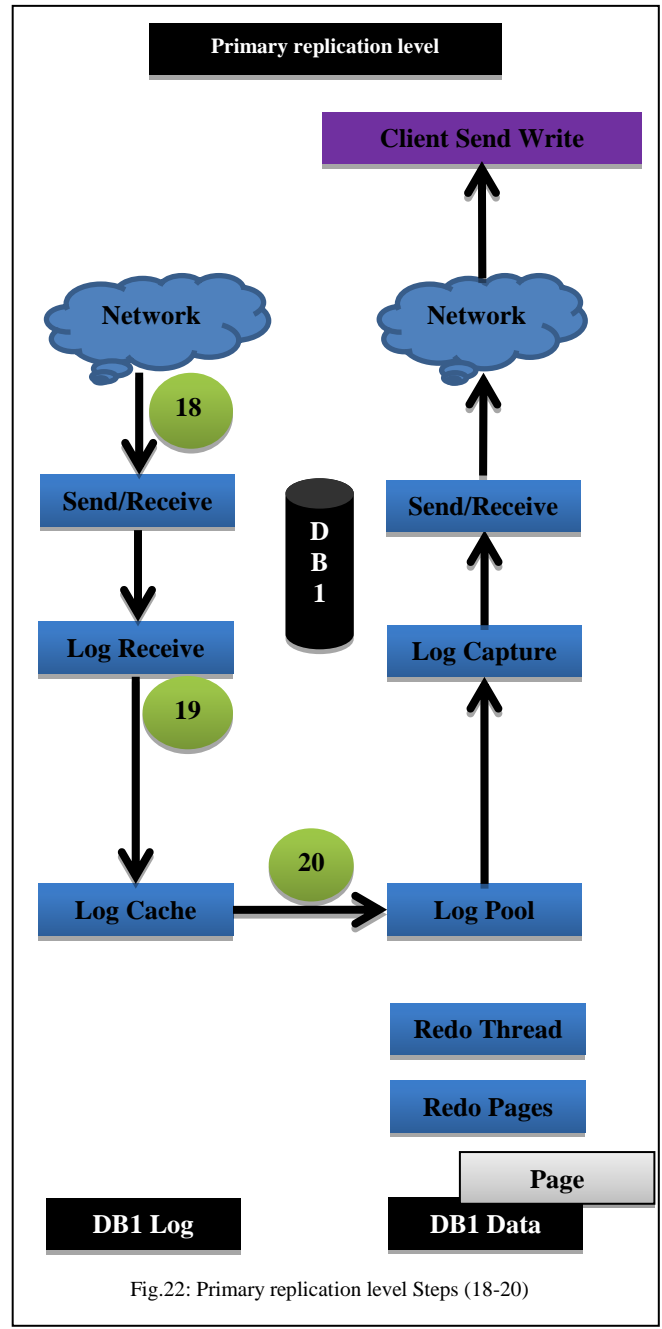

Fig.22: Primary replication level Steps (18-20)

```
Begin

Input:

Client_Write_Request

Secondary_Write_Log_sequence_number=Nothing

Status= acknowledgement from synchronous replication Level

Primary_Write_log_sequence_number= P01W0001

If server= primary and

Client_Write_Request<> Nothing and

Primary_Write_log_sequence_number<>Nothing

Secondary_Write_Log_sequence_number=Nothing and

Status= acknowledgement from synchronous replication Level Then

//The Primary finishes committing the data

Primary_Finish_Commit_Write_Request ()

End if

Display= acknowledge message is sent back to the clients

End
```
Fig.23 Pseudo code for phase three (the primary receive acknowledgement the write request from Secondary in synchronous replication level and send acknowledge to client)

Phase four: the Secondary in semi synchronous replication level receive primary commit write request from Secondary in synchronous replication level:

Fig.24 presents the processes in details that happened in the Secondary in semi synchronous replication level after primary commit is come back from the Secondary in synchronous replication level and in the Secondary in semi synchronous replication level dose not responsible for sending acknowledge back to the client, as shown in the pseudo code in Fig.25.

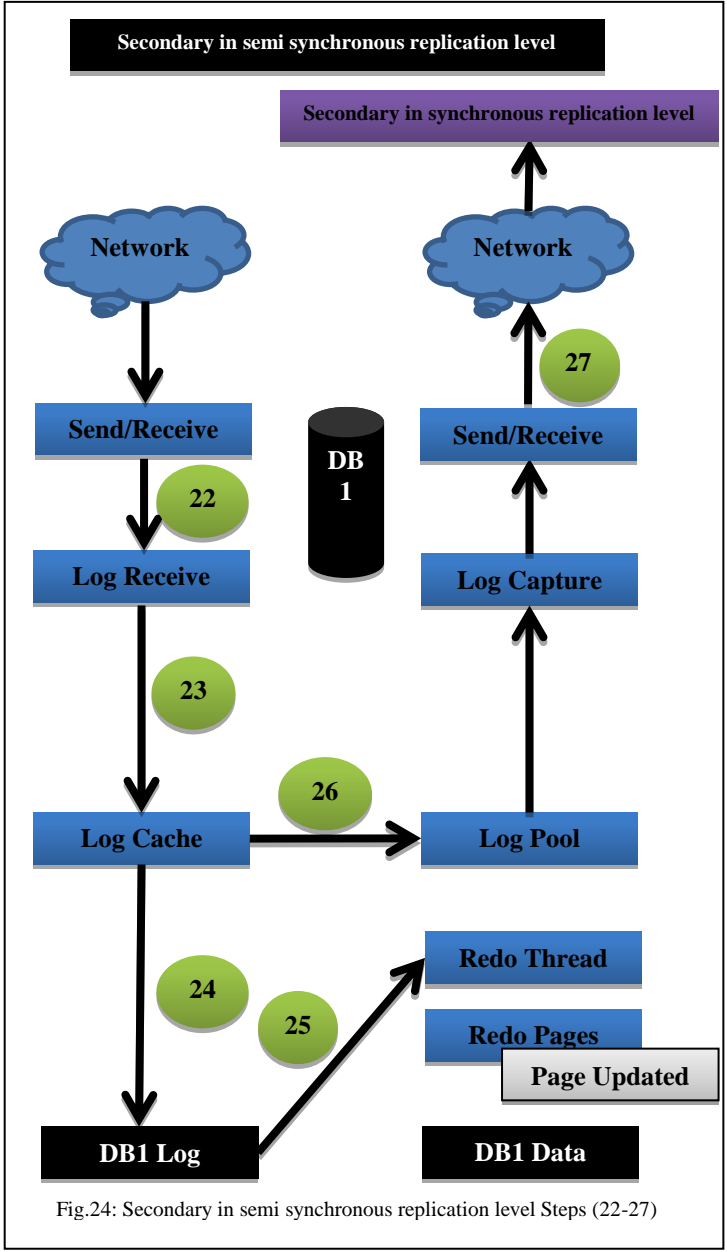

Fig.24: Secondary in semi synchronous replication level Steps (22-27)

```
Begin
Input:
Client_Write_Request
Secondary_Write_Log_sequence_number=Nothing
Status= primary commit
Primary_Write_log_sequence_number= P01W0001
If server= secondary in semi synchronous replication and
Client_Write_Request<> Nothing and
Primary_Write_log_sequence_number<>Nothing
Secondary_Write_Log_sequence_number=Nothing and
Status= primary commit Then
//write the log to the database log and database data
Applying_Write_Request ()
/* Determine_Status is a Function that determine status of the transaction for each phase*/
Set Status= Determine_Status (Client_Write_Request)
/*
Send_Staus_acknowledgement_synchronous_replication_Level () is a function that send acknowledgement to the secondary server in synchronous replication Level */
Send_Staus_acknowledgement_synchronous_replication_Level ()
End if
Outputs:
Client_Write_Request
Secondary_Write_Log_sequence_number=Nothing
Status= acknowledgement from semi synchronous replication Level
Primary_Write_log_sequence_number=P01W0001
End
```

Fig.25 Pseudo code for phase four (the Secondary in semi synchronous replication level receive primary commit write request from Secondary in

## 2.6 Failure cause and Solutions

An automatic failover occurs in response to a failure and we know that a failure happened if any server on the three levels dose not response to the other server at the session-timeout period so the connected servers is decided that the server is down as in the following automatic failover situations:

A) If Secondary in semi synchronous replication level consider failure because the session-timeout period is end. All connection will be transfer to the Secondary in synchronous replication level.

B) If Secondary in synchronous replication level consider failure because the session-timeout period is end. All connection will be transfer to the Primary server and we return to work to old criteria so all secondary servers in semi synchronous replication level will be open just for read only until the related secondary server in synchronous replication level is up again.

C) If Primary consider failure because the session-timeout period is end. One of the Secondary in synchronous replication level will be the new primary as the priority number that assign to each one of the Secondary in synchronous replication level.

## 2.7 Conclusion

In this paper presents a new model for database replication in private cloud availability regions. Using this model, each server in the region will be open for read/write request. Therefore using the new model the data will be transfer to all servers in the region synchronously and semi synchronously.

Since each server in the new model on the whole region have a copies of data, And the performance is improved through reduced latency, and this happened by letting users access nearby replicas and avoiding remote network access, and through increased throughput, by letting multiple computers serve the data simultaneously.

## 2.8 Future Work

This research has been pointed out the following recommendations for the Future Work.
1. Implements the proposed model for database replication in private cloud availability regions.
2. Evaluation and testing the new model for database replication in private cloud availability regions.

**References**

[1] R. Shipsey, (2010), Information systems: foundations of e-business, Volume 2, Department of Computing, Goldsmiths College, University of London.
[2] Hitachi Data Systems (2007). Synchronous Data Replication.
[3] Ashok,G., Randal,P.S. (2008), SQL Server Replication: Providing High Availability using Database Mirroring, Microsoft.
[4] IBM, (2010). Cloud computing insights from110 implementation projects.
[5] Mell, P. & Grance, T. (2011).The NIST Definition of Cloud Computing.USA: National Institute of Standards and Technology.
[6] Newton, J. (2010). The Ethics and Security of Cloud Computing. London: International Legal Technology Association.
[7] Saito, Y. & Shapiro, M. (2005). Optimistic Replication.ACM Computing Surveys, Vol. V, No. N, 3 2005.
[8] Amazon (2011).Amazon Web Services: Overview of Security Processes, Amazon.
[9] Abadi, D.J. (2009), Data Management in the Cloud: Limitations and Opportunities, IEEE.
[10] Kemme, B. & Alonso, G. (2010). Database Replication: a Tale of Research across Communities. Proceedings of the VLDB Endowment, Vol. 3, No. 1.

Ala'a Atallah A. Al-Mughrabi: graduate with B.Sc. in Computer Science from Petra University, Amman-Jordan in 2005, I'm Microsoft Certified Professional Developer (MCPD) on Microsoft Visual Studio 2008, Microsoft Visual Studio 2010 for Windows and Web Applications. I have 8 years' experience in Development Field. This paper is a summary of my thesis for master degree in Computer Information System in Middle East University, Computer Information System Dept.

Dr. Hussein H. Owaied: graduate with B.Sc. in mathematics from Al Mustansiryiah University, Baghdad-IRAQ in 1977, Postgraduate Diploma in Computer Science from UMIST, U.K, in 1984, PhD in Computer Science from Bradford University, U.K. in 1988. Currently is Associate Professor at Middle East University, Dept. of Computer Science. Usually the courses of Artificial Intelligence and Knowledge-Based Systems are the most of interesting for me. But to complete my duties I taught many courses such as Operating System, Logic Design, Distributed Information Systems, Coding and Information Theory, PROLOG, PASCAL, C++, Data Structures, or Computer Architecture. I have 24 years experience teaching most of the courses in computer science for different departments at many Universities for different levels (Undergraduate and Postgraduate) studies.